Theory of Refraction, Ray-Wave Tilt, Hidden Momentum, and Apparent Topological Phases in Isotropy-Broken Materials based on Electromagnetism of Moving Media

Maxim Durach (mdurach@georgiasouthern.edu)

Center for Advanced Materials Science, Department of Biochemistry, Chemistry & Physics, Georgia Southern University, Statesboro, GA 30460, USA

One of the problems of physics arguably greater in stature than even mathematical Hilbert's problems is the mysterious nature of electromagnetic momentum in materials. In this paper we show that the difference between the Minkowski and Abraham momenta, which is composed of the Roentgen and Shockley hidden momenta, is directly related to the phenomenon of refraction and the tilt of rays from the wavefront propagation direction. We demonstrate that individual electromagnetic waves with non-unit indices of refraction $n$ appear as quasistatic high-k waves to an observer in the proper frames of the waves. When Lorentz transformed into the material rest frames these high-k waves are Fresnel-Fizeau dragged from rest to their phase velocities and acquire longitudinal hidden momentum and related refractive properties. On the material level all electromagnetic waves belong to Fresnel wave surfaces topologically classified according to hyperbolic phases by Durach and determined from the electromagnetic material parameters. To moving observers, material parameters appear modified, which leads not only to the alterations of Fresnel wave surfaces, but even the topological classes of the materials may appear differently in moving frames. We discuss the phenomenon of the electromagnetic momentum tilt, defined as non-zero angle between Abraham and Minkowski momenta or equivalently between the rays and the wavefront propagation direction. We show that momentum tilt is only possible in isotropy-broken media, where **E** and **H** fields can be longitudinally polarized in presence of electric and magnetic bound charge waves. The momentum tilt can be understood as differential aberration of rays and waves when observed in material rest frame.

1. Introduction

The make up of light has captivated the humanity since Biblical times [1,2]. This continued in the scholarly works of Greco-Roman [3-5] and Islamic worlds [6-8]. In modern electromagnetism the generic isotropy is described by bi-isotropic diagonal tensors of dielectric permittivity $\hat{\epsilon} = \epsilon \hat{1}$, magnetic permeability $\hat{\mu} = \mu \hat{1}$ and magnetoelectric couplings $\hat{X}, \hat{Y}$ [9]. For isotropic materials the concepts of optical rays and electromagnetic waves progressed in the works of Pierre de Fermat in 1662 and Christiaan Huygens in 1678 [10], to produce the understanding that in isotropic media rays are directed perpendicular to the wavefronts. Furthermore, the polarization of light was established by Étienne-Louis Malus in 1811 [11] and clarified by Augustin-Jean Fresnel in 1821 [12] as being transverse to the ray and wavefront propagation direction in isotropic media. The magnitudes of k-vectors $k = k_0 n$ of all electromagnetic waves in isotropic media are independent of propagation direction, which allows us to consider indices of refraction $n$ of isotropic media, such as water and glass, as material parameters.

The observations of isotropy breaking in electromagnetic materials has been recorded since the 1669, when the double refraction by Iceland spar was reported by Rasmus Bartholin [13]. Understanding of this phenomenon grew through the works of Christiaan Huygens and Issac Newton [14] and culminated in the development of the concept of Fresnel wave surface $\mathcal{H}(\boldsymbol{k}, k_0) = 0$ and the optics of crystals by Augustin-Jean Fresnel in 1822 [15]. This work was supported by the prediction of conical refraction by William Rowan Hamilton in 1832 [16], who discovered it while developing his Hamiltonian geometrical optics [17]. In 1845 the Faraday rotation effect was discovered by Michael Faraday and gave rise to the studies of gyroelectromagnetic materials [18].

Electromagnetism of moving media has a tremendous impact on modern science. Fresnel-Fizeau drag, aberration of light, moving magnet and conductor problem, and negative aether drift tests formed the basis of Einstein's development of the theory of relativity [19]. In 1888 Wilhelm Conrad Roentgen discovered that a dielectric moving through an electric field creates magnetic field - the first observation of bianisotropy in moving media in the form of Roentgen interaction, i.e. Roentgen hidden momentum [20]. In 1905 Harold Albert Wilson demonstrated the electrical polarization of a dielectric, moving in magnetic field, which was later associated with Shockley hidden momentum [21]. It has been demonstrated that even isotropic media appear bianisotropic to moving observers [22]. Stationary bianisotropic crystals were first studied by Landau, Lifshitz, and Dzyaloshinskii in 1957-1959 [23,24]. For several decades now the field of bianisotropics and metamaterials occupies the central role in optics [25-29], with refraction in both isotropic [30,31] and isotropy-broken media [32,33] being one of the foci of research.

Fresnel wave surfaces of generic bianisotropic materials with arbitrary material parameters are quartic surfaces in k-space and are described by Tamm-Rubilar tensors $T_{ijlm}$ [34-36]

$$\mathcal{H}(\boldsymbol{k}, k_0) = \sum_{i+j+l+m=4} [T_{ijlm} k_x^i k_y^j k_z^l k_0^m] = 0 \qquad (1)$$

Topological asymptotic skeletons of the iso-frequency surfaces [Eq. (1)] can be found in the high-k limit $k \gg k_0$. The quasistatic high-k waves in materials tend to the conical surfaces given by Durach high-k characteristic function [32,33,37]

$$h(\boldsymbol{k}) = \mathcal{H}(k \to \infty, k_0) = \sum_{i+j+l=4} [T_{ijl0} k_x^i k_y^j k_z^l] = (\boldsymbol{k}^T \hat{\epsilon} \boldsymbol{k})(\boldsymbol{k}^T \hat{\mu} \boldsymbol{k}) - (\boldsymbol{k}^T \hat{X} \boldsymbol{k})(\boldsymbol{k}^T \hat{Y} \boldsymbol{k}) = 0 \quad (2)$$

By investigating the properties of Eq. (2) Durach et al. [32,33] established that all optical materials can be topologically classified using 5 hyperbolic classes: non-, mono-, bi-, tri-, and tetra-hyperbolic materials. The prefix in the name of the class indicates the number of double cones in high-k limit in the Fresnel wave surface. Rays and waves in the media with broken isotropy are characterized by ray and wave surfaces corresponding to non-parallel ray vectors $\boldsymbol{s}$ and wave vectors $\boldsymbol{k}$ [23]. In Hamiltonian geometrical optics this is expressed by one of the pair of Hamilton equations for the wave vector $\boldsymbol{k}$ and the ray vector $\boldsymbol{s}$ of electromagnetic field [38]

$$\frac{d\boldsymbol{r}}{d\tau} = \frac{\partial \mathcal{H}(\boldsymbol{k},k_0)}{\partial \boldsymbol{k}} = \boldsymbol{s} \qquad (3)$$

where $\tau$ is a parameter proportional to the arclength along the ray. This signifies that for the electromagnetic fields the canonical momentum is directed along the wave vector $\boldsymbol{k}$, while the kinetic momentum is directed along the ray vector $\boldsymbol{s}$, which in accordance with Eq. (3) is normal to the Fresnel wave surface [23]. The ray-wave duality principle, introduced by Fedor I. Fedorov, states the existence of dual media symmetric upon interchange between ray and wave vectors $\boldsymbol{s} \leftrightarrow \boldsymbol{k}$ [39,40].

The idea that the electromagnetic fields carry linear momentum as they propagate and exert pressure was introduced by James Clerk Maxwell in 1862 [41]. The pressure of light was first measured by Peter Nikolaevich Lebedew in 1899, which became the first quantitative confirmation of Maxwell's theory of electromagnetism [42]. This, however, was followed by already a century-long Abraham-Minkowski controversy about the proper definition of the electromagnetic momentum volume density with two different proposals by Hermann Minkowski in 1908 [43] $\boldsymbol{g}_{Min} = \frac{1}{4\pi c} \boldsymbol{D} \times \boldsymbol{B}$ and Max Abraham in 1909 [44] $\boldsymbol{g}_{Abr} = \frac{1}{4\pi c} \boldsymbol{E} \times \boldsymbol{H}$. Abraham's definition of momentum is proportional to the Poynting vector $\boldsymbol{S} = \frac{c}{4\pi} \boldsymbol{E} \times \boldsymbol{H}$ describing the electromagnetic energy flux density and is directed along $\boldsymbol{s}$. Minkowski's momentum is directed along $\boldsymbol{k}$ in source-free regions. Inside isotropic media the Abraham and Minkowski momentum densities are directed along wave propagation and have different magnitudes $g_{Min} = g_0 n$ and $g_{Abr} = g_0/n$. The resolution of Abraham-Minkowski controversy for an isotropic dielectric medium was proposed in 2010 by Barnett [45] who attributed the Minkowski momentum to the canonical momentum of electromagnetic field and the Abraham momentum to the kinetic momentum of the field (see also Eq. (3)), and the difference between them to the Roentgen interaction, which corresponds to the Roentgen hidden momentum with density $\boldsymbol{g}_{RH} = \frac{1}{c} \boldsymbol{P} \times \boldsymbol{B}$ [20,46]. Please note, however, that the extensive Abraham-Minkowski controversy literature is focused exclusively on isotropic media in which $n$ is a material parameter, and no investigation has been done on Abraham-Minkowski controversy in isotropy-broken media, where $n$ is not a material parameter [32,33,37].

In a parallel debate multiple definitions of the electromagnetic force applied to a medium were proposed with the discussion mainly revolving around the Lorentz force and the Einstein-Laub force, culminating in the formulation of the Mansuripur's paradox [47]. The resolution of the Einstein-Laub-Lorentz controversy was presented in 2017 by Durach [48] who proposed an expression for the force applied to an arbitrary dielectric medium, including dispersive and isotropy-broken media, which corresponds to the Lorentz force when no spin polarization of electrons is induced and to the Eistein-Laub force otherwise, while the difference between them was attributed to absorption of the spin angular momentum of light by media through spin forces. In plasmonic metals the spin forces lead to the pinning of the plasmon drag effect (PLDE) forces to the angstrom-thick surface layer as predicted by Durach [48] and later experimentally confirmed by surface sensitivity measurements of PLDE at NIST in 2019 [49].

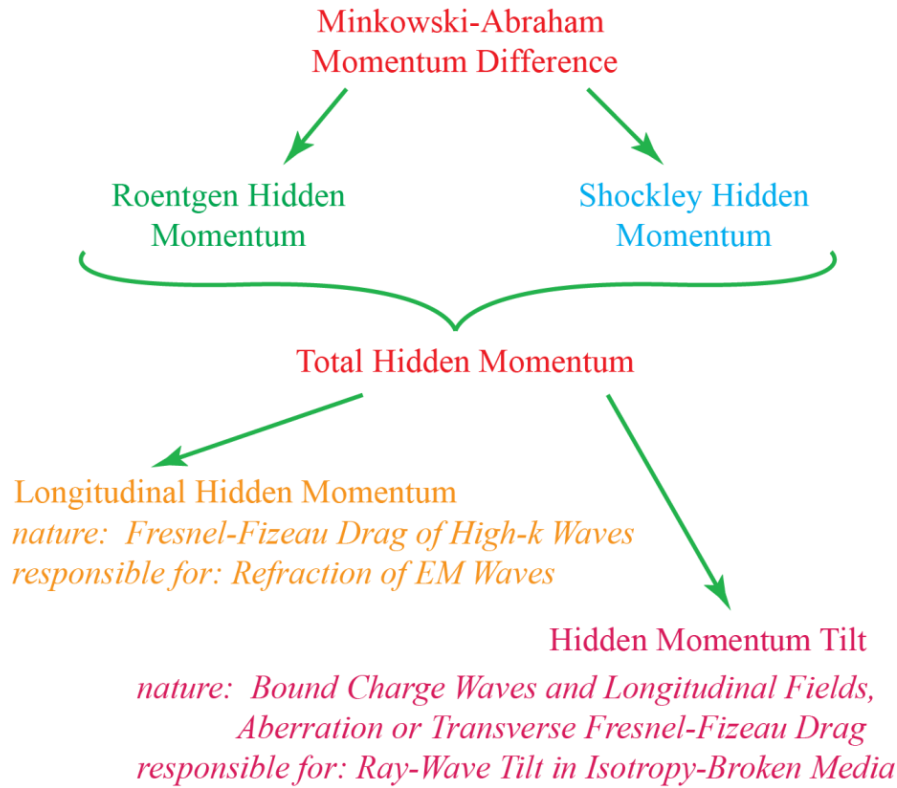

*Fig. 1. Summary of the terminology used here, and an outline of some of the findings presented in this paper.*

Note that both Abraham-Minkowski controversy and Mansuripur's paradox are related to the concept of Shockley hidden momentum with density $\boldsymbol{g}_{SH} = \frac{1}{c}\boldsymbol{E} \times \boldsymbol{M}$ introduced in 1961 [50, 51]. Both Abraham-Minkowski and Einstein-Laub-Lorentz problems are aggravated by the plethora of different definitions of electromagnetic momenta and forces, which are based on different ways of structuring Poynting theorems and electromagnetic energy-momentum tensors, the number of which may range from 4 to 729 according to different accounts [52,53]. The deeper understanding of electromagnetic momentum and its transfer in media is far from a glorified purely academic puzzle, but has huge practical implications ranging from solar sails and comet tails [54] to optical tweezers and wrench devices [55], to hyperlenses [56] and PLDE sensors [48], etc. In Fig. 1 we outline the summary of the terminology we use in this paper and some of the findings we present here.

In this paper we show that the hidden momentum in an electromagnetic wave is directly related to its refractive index and is acquired by the wave, when transformed from its proper frame into the material rest frame. We demonstrate that isotropy breaking in electromagnetic materials induces bound charge waves and non-transverse polarization of electromagnetic waves. This is directly related to the difference between ray vectors $s$ and wave vectors $k$ directions, or equivalently to the difference in the Minkowski and Abraham momentum directions and can be understood as differential aberration of rays and waves when observed in material rest frames, while those frames are in relative motion with respect to the frames in which ray and wave sources coincide.

2. The hidden momentum, refraction, ray-wave tilt, and bound charge waves

Based on the definitions of the electric and magnetic inductions $D = E + 4\pi P$ and $B = H + 4\pi M$ the volumetric density of the total hidden momentum $\Delta g$, which is the difference between the Abraham momentum $g_{Abr}$ and Minkowski momentum $g_{Min}$, is equal to the sum of the Roentgen hidden momentum $g_{RH}$ and Shockley hidden momentum $g_{SH}$ (see the scheme in Fig. 1):

$$\Delta g = g_{Abr} - g_{Min} = \frac{1}{4\pi c} D \times B - \frac{1}{4\pi c} E \times H = g_{RH} + g_{SH} = \frac{1}{c} P \times B + \frac{1}{c} E \times M$$

We consider a generic linear material with the most general bianisotropic constitutive relations:

$$\begin{pmatrix} D \\ B \end{pmatrix} = \widehat{M} \begin{pmatrix} E \\ H \end{pmatrix} = \begin{pmatrix} \hat{\epsilon} & \hat{X} \\ \hat{Y} & \hat{\mu} \end{pmatrix} \begin{pmatrix} E \\ H \end{pmatrix}, \tag{4}$$

A plane wave with wave vector $k = (0,0,k_z)$ propagating through a material described by Eq. (4) is carrying fields $E, H, D, B \propto e^{i(kr-\omega t)}$. The time-averaged longitudinal component of the hidden momentum is

$$\overline{\Delta g_z} = \frac{1}{4\pi c} \hat{z} \cdot \text{Re}\{D^* \times B - E^* \times H\} \tag{5}$$

We utilize the following identities that follow from Maxwell's equations:

$$\text{Re}\{D^* \times B\} = \frac{1}{k_0^2} (k \cdot \text{Re}\{E \times H^*\}) k \tag{6}$$

$$\frac{k_0}{2} \text{Re}\{E^* \cdot D + H^* \cdot B\} = (k \cdot \text{Re}\{E^* \times H\}) \tag{7}$$

Normalizing the electromagnetic fields such that $\frac{1}{8\pi}\text{Re}\{\boldsymbol{E}^* \cdot \boldsymbol{D} + \boldsymbol{H}^* \cdot \boldsymbol{B}\} = U$, where $U$ is the energy density of electromagnetic field, we arrive at

$$\frac{c\overline{\Delta g_z}}{U} = nf = n\left(1 - \frac{1}{n^2}\right), \tag{8}$$

where the index of refraction is $n = \boldsymbol{k}/k_0$, the phase velocity is $v_{ph} = c/n$.

From Eq. (8) we see that the index of refraction of an electromagnetic wave is directly related to the longitudinal component of the hidden momentum $\overline{\Delta g_z}$ in the wave and can be expressed as

$$n = \sqrt{1 + \frac{1}{4}\left(\frac{c\overline{\Delta g_z}}{U}\right)^2} + \frac{1}{2}\left(\frac{c\overline{\Delta g_z}}{U}\right) \tag{9}$$

This result provides a closed-form direct relationship between the index of refraction of the waves and the amplitudes of the waves $\boldsymbol{E}, \boldsymbol{H}, \boldsymbol{D}, \boldsymbol{B}$.

Let us turn to the transverse component of the hidden momentum $\boldsymbol{\Delta g}_\perp$. According to the index of refraction operator method [32,33] the longitudinal fields in the wave can be expressed as

$$(E_z, H_z)^T = -\widehat{M}_z^{-1} \cdot \widehat{M}_{z\|} \cdot (E_x, E_y, H_x, H_y)^T \tag{10}$$

where

$$\widehat{M}_{z,\|} = \begin{pmatrix} \epsilon_{31} & \epsilon_{32} & X_{31} & X_{32} \\ Y_{31} & Y_{32} & \mu_{31} & \mu_{32} \end{pmatrix}, \quad \widehat{M}_{z,z} = \begin{pmatrix} \epsilon_{33} & X_{33} \\ Y_{33} & \mu_{33} \end{pmatrix} \tag{11}$$

In isotropic media $\widehat{M}_{z,\|} = \widehat{0}$, and all fields are solenoidal with purely transverse amplitudes. The isotropy breaking leads to non-zero $\widehat{M}_{z,\|}$ and appearance of longitudinal components $(\boldsymbol{k} \cdot \boldsymbol{E}), (\boldsymbol{k} \cdot \boldsymbol{H}) \neq 0$ and divergences $(\boldsymbol{\nabla} \cdot \boldsymbol{E}), (\boldsymbol{\nabla} \cdot \boldsymbol{H}) \neq 0$. In source-free regions the induction fields $\boldsymbol{D}, \boldsymbol{B}$ do not have divergence $(\boldsymbol{\nabla} \cdot \boldsymbol{D}), (\boldsymbol{\nabla} \cdot \boldsymbol{B}) = 0$, which means that waves in isotropy-broken materials carry effective bound electric and magnetic charge waves

$$\rho_{be} = -\boldsymbol{\nabla} \cdot \boldsymbol{P} = \frac{1}{4\pi}\boldsymbol{\nabla} \cdot \boldsymbol{E} = \frac{1}{4\pi}(i\boldsymbol{k} \cdot \boldsymbol{E})e^{i(\boldsymbol{kr}-\omega t)} = -(i\boldsymbol{k} \cdot \boldsymbol{P})e^{i(\boldsymbol{kr}-\omega t)} \tag{12}$$

$$\rho_{me} = -\boldsymbol{\nabla} \cdot \boldsymbol{M} = \frac{1}{4\pi}\boldsymbol{\nabla} \cdot \boldsymbol{H} = \frac{1}{4\pi}(i\boldsymbol{k} \cdot \boldsymbol{H})e^{i(\boldsymbol{kr}-\omega t)} = -(i\boldsymbol{k} \cdot \boldsymbol{M})e^{i(\boldsymbol{kr}-\omega t)} \tag{13}$$

Since the fields $\boldsymbol{D}$ and $\boldsymbol{B}$ are solenoidal and are transverse to the phase propagation direction $\boldsymbol{k}$

$$\overline{\boldsymbol{\Delta g}_\perp} = \frac{1}{c}\,\text{Re}\{P_z^*\widehat{\boldsymbol{k}} \times \boldsymbol{B} + \boldsymbol{D}^* \times M_z\widehat{\boldsymbol{k}} - 4\pi P_z^*\widehat{\boldsymbol{k}} \times \boldsymbol{M}_\| - 4\pi \boldsymbol{P}_\|^* \times M_z\widehat{\boldsymbol{k}}\}$$

$$= \frac{1}{c}\left(\text{Re}\{P_z\widehat{\boldsymbol{k}} \times \boldsymbol{H}_\|^* + \boldsymbol{E}_\|^* \times M_z\widehat{\boldsymbol{k}}\}\right) = -\frac{k_0}{ck^2}\left(\text{Re}\{i\rho_{be}\boldsymbol{D}^* + i\rho_{bm}\boldsymbol{B}^*\}\right) \tag{14}$$

The transverse component of the hidden momentum density $\overline{\Delta g_\perp}$ is responsible for the ray-wave tilt in isotropy-broken media. The tilt appears because the Abraham momentum density $g_{Abr}$ is directed along the ray vector $s$, while Minkowski momentum density $g_{Min}$ is directed along the wave vector $k$. We see from Eq. (14) that the difference between directions of $g_{Abr}$ and $g_{Min}$ is due to the existence of the transverse component of the hidden momentum $\overline{\Delta g_\perp}$, which requires non-zero longitudinal polarization $P_z$ and magnetization $M_z$ of the material. According to Eq. (10) the longitudinal polarization and magnetization are only possible in isotropy-broken media and are related to the propagation of electric and magnetic bound charge waves described by Eqs. (12)-(13).

3. Topological phases of media in moving frames

The presence of the Fizeau-Fresnel dragging coefficient $f = \left(1 - \frac{1}{n^2}\right)$ in Eq. (8) is intriguing and merits further investigation in this manuscript. To gain better understanding, we consider Fresnel wave surfaces of materials in moving frames. We look at the Lorentz transformations into a frame $S_\beta$ moving with velocity $V = c\beta$ from the material rest frame $S_{mat} = S_{\beta=0}$. We express the material relations in the material rest frame $S_{mat}$ in $EH$ representation as

$$\begin{pmatrix} D \\ B \end{pmatrix}_{\beta=0} = \widehat{M}_{\beta=0} \begin{pmatrix} E \\ H \end{pmatrix}_{\beta=0} = \begin{pmatrix} \hat{\epsilon} & \hat{X} \\ \hat{Y} & \hat{\mu} \end{pmatrix} \begin{pmatrix} E \\ H \end{pmatrix}_{\beta=0},$$

This can be converted into the Lorentz covariant $EB$ representation of Kong [57], since both $EB$ and $DH$ transform from $S_{mat}$ into $S_\beta$ as

$$\begin{pmatrix} D \\ H \end{pmatrix}_\beta = \hat{L} \begin{pmatrix} D \\ H \end{pmatrix}_{\beta=0} = \gamma \begin{pmatrix} \hat{\alpha}^{-1} & \hat{\beta} \\ -\hat{\beta} & \hat{\alpha}^{-1} \end{pmatrix} \begin{pmatrix} D \\ H \end{pmatrix}_{\beta=0} \text{ and } \begin{pmatrix} E \\ B \end{pmatrix}_\beta = \hat{L} \begin{pmatrix} E \\ B \end{pmatrix}_{\beta=0},$$

$$\hat{\alpha}^{-1} = \hat{1} + \left(\frac{1}{\gamma} - 1\right)\frac{\beta\beta}{\beta^2}, \qquad \hat{\beta} = \beta \times$$

In $EB$ representation

$$\begin{pmatrix} D \\ H \end{pmatrix}_{\beta=0} = \hat{C}_{\beta=0} \begin{pmatrix} E \\ B \end{pmatrix}_{\beta=0} = \begin{pmatrix} \hat{C}_{DE} & \hat{C}_{DB} \\ \hat{C}_{HE} & \hat{C}_{HB} \end{pmatrix} \begin{pmatrix} E \\ B \end{pmatrix}_{\beta=0} = \begin{pmatrix} \hat{\epsilon} - \hat{X}\hat{\mu}^{-1}\hat{Y} & \hat{X}\hat{\mu}^{-1} \\ -\hat{\mu}^{-1}\hat{Y} & \hat{\mu}^{-1} \end{pmatrix} \begin{pmatrix} E \\ B \end{pmatrix}_{\beta=0}$$

$$\begin{pmatrix} D \\ H \end{pmatrix}_\beta = \hat{C}' \begin{pmatrix} E \\ B \end{pmatrix}_\beta = (\hat{L}\hat{C}_{\beta=0}\hat{L}^{-1}) \begin{pmatrix} E' \\ B' \end{pmatrix}$$

Returning to the $EH$ representation

$$\begin{pmatrix} D \\ B \end{pmatrix}_\beta = \widehat{M}_\beta \begin{pmatrix} E \\ H \end{pmatrix}_\beta = \begin{pmatrix} \left(\hat{C}'_{DE} - \hat{C}'_{DB}(\hat{C}'_{HB})^{-1}\hat{C}'_{HE}\right) & \hat{C}'_{DB}(\hat{C}'_{HB})^{-1} \\ -(\hat{C}'_{HB})^{-1}\hat{C}'_{HE} & (\hat{C}'_{HB})^{-1} \end{pmatrix} \begin{pmatrix} E \\ H \end{pmatrix}_\beta$$

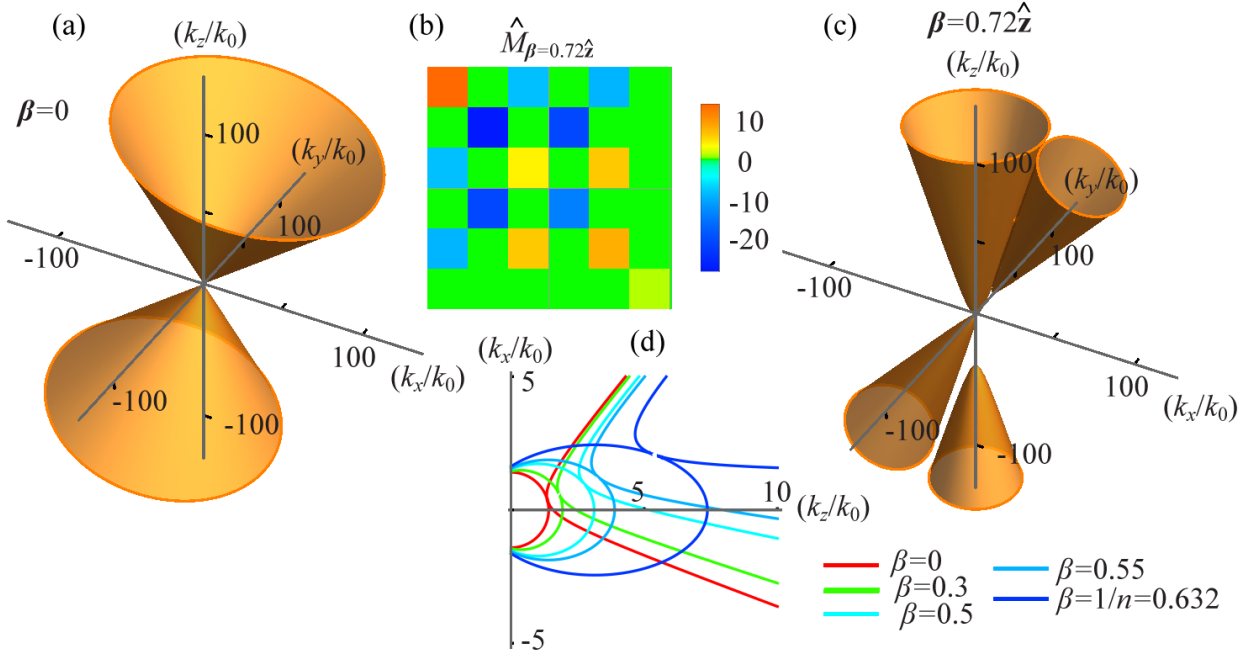

*Fig. 2. Topological transformation of material with $\hat{\epsilon} = \text{diag}\{2, 2-1\}$; (a) Fresnel wave surface $\mathcal{H} = 0$ in $S_{mat}$ (mono-hyperbolic phase). (b) Material parameter matrix $\widehat{M}_{\beta=0.72\hat{z}}$ in $S_{\beta=0.72\hat{z}}$. (c) Fresnel wave surface $\mathcal{H}_{\beta=0.72\hat{z}} = 0$ for an observer in $S_{\beta=0.72\hat{z}}$. (bi-hyperbolic class). (d) Cross-sections of Fresnel wave surfaces $\mathcal{H}_\beta = 0$ in $k_x k_z$-plane in different moving frames (color-coded to lower right of the panel).*

Since the material parameters are transformed from matrix $\widehat{M}_{\beta=0}$ in the frame $S_{mat}$ to $\widehat{M}_\beta$ in $S_\beta$, the Fresnel wave surfaces $\mathcal{H}_\beta = 0$ in the moving frame are also transformed. Indices of refraction of waves in the moving frame $S_\beta$ are found as eigenvalues of the index of refraction operator $\widehat{N}_\beta$ [32,33]. It is natural to expect that since material parameters are changed, not only the Fresnel wave surfaces are modified, but even the topological hyperbolic classes of the materials can appear differently in moving frames. Indeed, in Fig. 2 we show topological transformation of a nonmagnetic material with $\hat{\epsilon} = \text{diag}\{2, 2-1\}$ which is free from magnetoelectric coupling in moving frames. In Fig.2(a) we show the Fresnel wave surface $\mathcal{H} = 0$ in $S_{mat}$, which corresponds to a mono-hyperbolic material. In a moving frame $S_{\beta=0.72\hat{z}}$ the material parameters are described by matrix $\widehat{M}_{\beta=0.72\hat{z}}$, which is color coded in the panel Fig. 2(b). For an observer in $S_{\beta=0.72\hat{z}}$ the Fresnel wave surface $\mathcal{H}_{\beta=0.72\hat{z}} = 0$ is shown panel Fig. 2(c) and appears to be in the bi-hyperbolic class. In Fig. 2(d) we show modification of a cross-section of the Fresnel wave surface $\mathcal{H}_\beta = 0$ in $k_x k_z$-plane in different moving frames as color-coded to lower right of the panel.

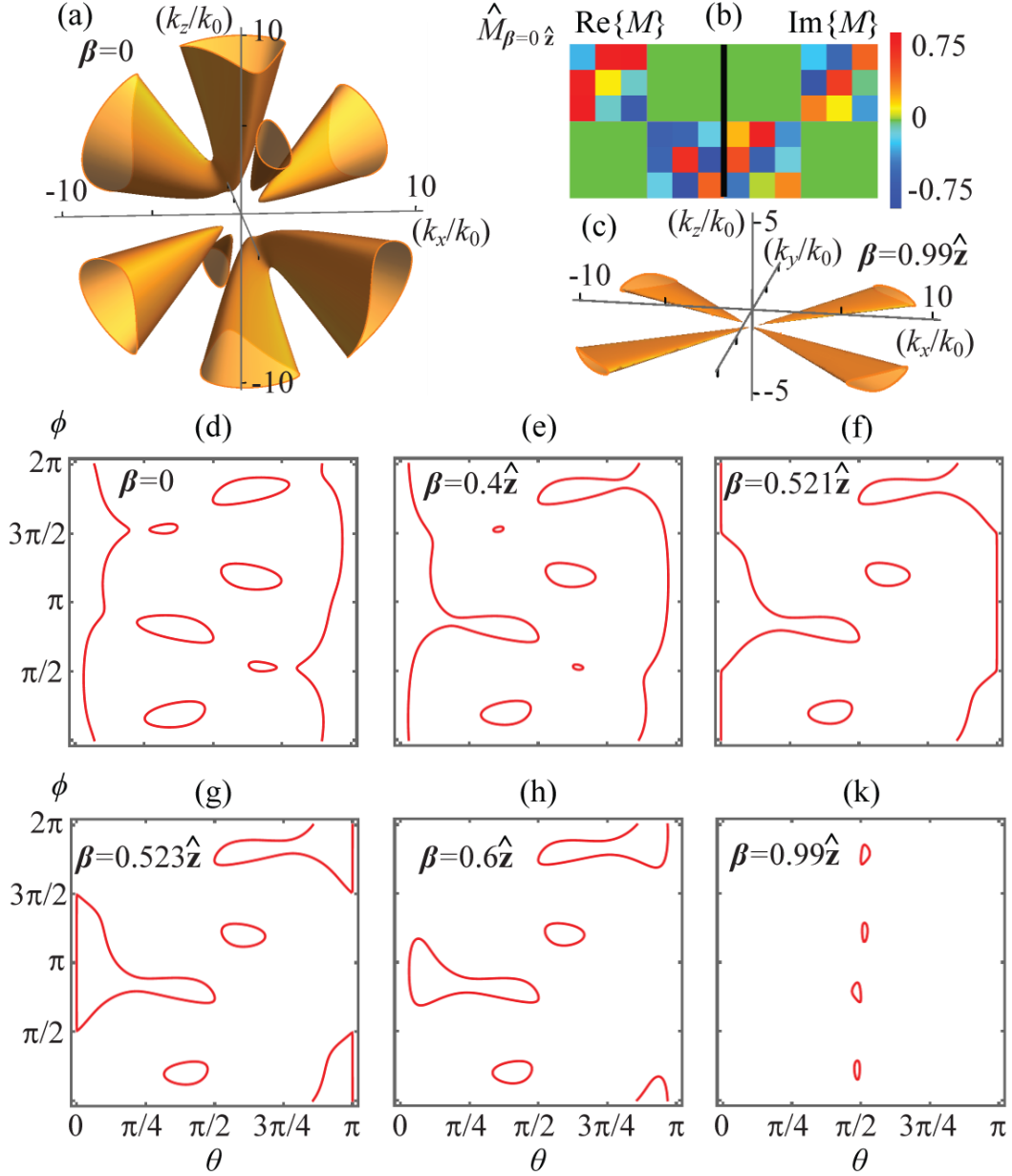

*Fig. 3. Topological transformation of a tetra-hyperbolic material; (a) Fresnel waves surface $\mathcal{H} = 0$ in the material rest frame $S_{mat}$ (tetra-hyperbolic class); (b) material parameters matrix $\hat{M}$ in $S_{mat}$; (c) Fresnel waves surface in the moving frame $S_{\boldsymbol{\beta}=0.99\hat{z}}$ (bi-hyperbolic material); (d)-(k) Topological phase transformations of Fresnel waves surfaces $\mathcal{H} = 0$ at high-k limit in moving frames from tetra-hyperbolic in $S_{mat}$ in panel (d), to tri-hyperbolic in panel (e), to bi-hyperbolic in frames (f)-(k).*

The topological classes can not only appear to increase to moving observers, but also to decrease. In Fig. 3 we consider a material with Fresnel waves surface $\mathcal{H} = 0$ shown in Fig. 3(a) and the material parameters matrix $\hat{M}$ in Fig. 3(b). It is a tetra-hyperbolic material in the material rest frame $S_{mat}$. However, in the moving frame $S_{\boldsymbol{\beta}=0.99\hat{z}}$ it is a bi-hyperbolic material, as shown in Fig. 3(c).

In panels (d)-(k) of Fig. 3 we show the Fresnel waves surfaces $\mathcal{H} = 0$ at their high-k limit $k \gg k_0$ and the transformations of the topological phases of the material in different moving frames from tetra-hyperbolic in $S_{mat}$ in panel (d), to tri-hyperbolic in panel (e), to bi-hyperbolic in frames (f)-(k).

4. Index of refraction in moving frames and Fresnel-Fizeau drag

To bring further insights into this consider the following. The index of refraction of water is 1.33. For most glasses it is around 1.5, etc. We assign indices of refraction to materials, but this is only valid for isotropic media, where all waves have the same index. This is no longer true in isotropy-broken media where different waves have different indices of refraction, and index of refraction is no longer a material parameter, but a property of the wave. The Lorentz transformations for the k-vector $\mathbf{k}$ and frequency $\omega = k_0 c$ from the material rest frame $S_{mat}$ into a frame $S_{\boldsymbol{\beta}}$ moving with velocity $\mathbf{V} = c\boldsymbol{\beta}$ are [57]

$$\mathbf{k}_{\boldsymbol{\beta}} = \mathbf{k} + \left\{(\gamma - 1)\frac{1}{\beta^2}(\boldsymbol{\beta} \cdot \mathbf{k}) - k_0\gamma\right\}\boldsymbol{\beta}$$

$$k_{0\boldsymbol{\beta}} = \gamma(k_0 - \boldsymbol{\beta} \cdot \mathbf{k})$$

Correspondingly, the Fresnel wave surface $\mathcal{H}(\mathbf{k}, k_0) = 0$ is modified to $\mathcal{H}_{\boldsymbol{\beta}}(\mathbf{k}_{\boldsymbol{\beta}}, k_{0\boldsymbol{\beta}}) = 0$ as shown in Figs. 2-3.

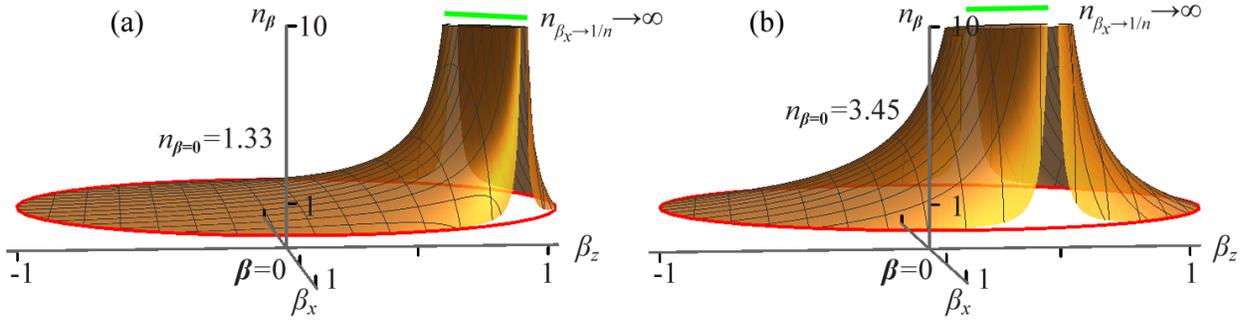

Fig. 4. Indices of refraction $n_{\boldsymbol{\beta}}$ as function of $\boldsymbol{\beta}$ for (a) $n_{\boldsymbol{\beta}=0} = 1.33$ and (b) $n_{\boldsymbol{\beta}=0} = 3.45$

More fundamentally, any electromagnetic wave with index of refraction $\mathbf{n} = \mathbf{k}/k_0$ acquires a new index of refraction $\mathbf{n}_{\boldsymbol{\beta}}$ in moving frames according to

$$\mathbf{n}_{\boldsymbol{\beta}} = \frac{\mathbf{k}_{\boldsymbol{\beta}}}{k_{0\boldsymbol{\beta}}} = \frac{\mathbf{n} + \left\{(\gamma - 1)\frac{1}{\beta^2}(\boldsymbol{\beta} \cdot \mathbf{n}) - \gamma\right\}\boldsymbol{\beta}}{\gamma(1 - \boldsymbol{\beta} \cdot \mathbf{n})} \qquad (15)$$

Note that no material parameters enter Eq. (15). This means that any wave with a certain index of refraction will appear to have its index of refraction changed in moving frames according to Eq.

(15), independently of the material medium that hosts this. For example, in Fig 4 we plot indices of refraction $n_\beta$ as seen in all possible moving frames as a function of $\boldsymbol{\beta}$ for two waves with $n_{\beta=0} = 1.33$ [Fig. 4(a)] and $n_{\beta=0} = 3.45$ [Fig. 4(b)] propagating the z-direction in the material rest frame $S_{mat}$.

As can be seen from Fig. 4, there are two universal features in the function given by Eq. (15). Firstly, for the frames moving with $\beta \to 1, \gamma \to \infty$ indicated by the red circles in Fig. 4 $\boldsymbol{n_\beta} \to -\boldsymbol{\widehat{\beta}}$. Note that in this case $k_\beta \to \infty$, but this does not result in a high-k waves, since $k_{0\beta} \to \infty$, meaning that in these frames $n_\beta \to 1$ and the waves appear vacuum-like as if they were under optical neutrality conditions [58]. We call these frames א-frames for אור - light in Hebrew or $S_\aleph$-frames [1,2]. Secondly, in frames moving with $\beta_x \to 1/n_{\beta=0}$, indicated by the green lines in Fig. 4, the index of refraction tends to infinity, since $k_{0\beta} \to 0$, and the waves appear as quasistatic high-k dark field. We call these frames ח-frames for חֹשֶׁךְ - darkness in Hebrew or $S_\text{ח}$-frames [1,2].

As we mentioned above, all waves in water or silicon have the same index of refraction 1.33 or 3.45 in material rest frame $S_{mat}$. Their indices transform according to Eq. (15) into moving frames. Nevertheless, the dot product $(\boldsymbol{\beta} \cdot \boldsymbol{n})$ is different for waves with different propagation direction, therefore they transform into waves with different $\boldsymbol{n_\beta}$, such that even isotropic material appears bianisotropic to a moving observer [22,57].

Our results for Fresnel wave surfaces, topological classes, and wave indices of refraction in moving frames are related to the Fresnel-Fizeau drag for the phase velocity of light. Consider a frame $S_{\beta\hat{z}}$ moving with respect to the material rest frame $S_{mat}$ with speed $V = \beta c$ in the direction of the wave propagation $\widehat{\boldsymbol{k}} = \hat{\boldsymbol{z}}$. In this case Eq. (15) turns into

$$\boldsymbol{n}_{\beta\hat{z}} = \frac{(n-\beta)}{1-\beta n}\hat{\boldsymbol{z}}$$

Correspondingly, the difference between the phase velocity $v_{ph,\beta}$ in the moving frame $S_{\beta\hat{z}}$ and $v_{ph}$ in the material rest frame $S_{mat}$

$$\frac{c}{n_{\beta\hat{z}}} = \frac{c}{n}\frac{(1-\beta n)}{\left(1-\frac{\beta}{n}\right)} \underset{\beta \ll 1/n}{\approx} \frac{c}{n} - c\beta\left(1 - \frac{1}{n^2}\right) = \frac{c}{n} - Vf \qquad (16)$$

$$\Delta v_{ph} = \frac{c}{n_{\beta\hat{z}}} - \frac{c}{n} = -c\beta\frac{1-\frac{1}{n^2}}{1-\frac{\beta}{n}} \underset{\beta \ll 1/n}{\approx} -Vf$$

This change in phase speed of light under frame transformation is called Fresnel-Fizeau drag with drag coefficient $f$.

5. Longitudinal hidden momentum and high-k waves in the proper darkness frames

Transformation of the Fresnel wave surfaces in moving frames, not only changes the indices of refraction of the waves, but also the normal directions to the Fresnel wave surfaces, which correspond to the ray propagation directions. The normals to the Fresnel wave surfaces are directed along the Poynting vectors and Abraham momentum [23]. To study this, we turn to the Lorentz transformations for the fields $E, H, D, B$ from the material rest frame $S_{mat}$ into a frame $S_\beta$ [57]. We obtain the transformations for the longitudinal components of the time-averaged Minkowski and Abraham momentum densities $\bar{g}_{Min}, \bar{g}_{Abr}$ and the field energy density $\bar{U}$

$$\bar{g}^{\beta\hat{z}}_{Min_z} = \gamma^2 \left( \bar{g}_{Min_z} - 2\frac{\bar{U}}{c}\beta + \bar{g}_{Abr_z}\beta^2 \right)$$

$$\bar{g}^{\beta\hat{z}}_{Abr_z} = \gamma^2 \left( \bar{g}_{Abr_z} - 2\frac{\bar{U}}{c}\beta + \bar{g}_{Min_z}\beta^2 \right)$$

$$\bar{U}_{\beta\hat{z}} = \gamma^2 \left( (1+\beta^2)\bar{U} - \beta c (\bar{g}_{Min_z} + \bar{g}_{Abr_z}) \right)$$

Correspondingly, the normalized hidden momentum density transforms as

$$\frac{c\overline{\Delta g}^{\beta\hat{z}}_z}{\bar{U}_{\beta\hat{z}}} = \frac{(1-\beta^2)\cdot c\overline{\Delta g}_z}{(1+\beta^2)U - \beta c (\bar{g}_{Min_z} + \bar{g}_{Abr_z})} \tag{17}$$

Consider electromagnetic waves in their proper frames, i.e. frames which are traveling with the phase speed of the wave $V = v_{ph} = c/n$ in the wave propagation direction with respect to the material rest frame $S_{mat}$, such that $\boldsymbol{\beta} = \boldsymbol{V}/c = \hat{\boldsymbol{z}}/n$. This is the darkness ⊓-frames $S_⊓$ we introduced above. In $S_⊓$-frames the waves become high-k waves $k_\beta \gg k_{0\beta}$, due to the Doppler effect $k_{0\beta} = \gamma(k_0 - \boldsymbol{\beta}\cdot\boldsymbol{k}) \to 0$ and $n_\beta \to \infty$. High-k waves are quasistatic $v_{ph,\beta} \to 0$ and according to Refs. [32,33,37] have purely longitudinal electric and magnetic fields $\boldsymbol{E}^\beta = E_z^\beta \hat{\boldsymbol{z}} = -4\pi P_z^\beta \hat{\boldsymbol{z}}$, $\boldsymbol{H}^\beta = H_z^\beta \hat{\boldsymbol{z}} = -4\pi M_z^\beta \hat{\boldsymbol{z}}$, as well as transverse fields $\boldsymbol{D}^\beta$ and $\boldsymbol{B}^\beta$, which are related to the longitudinal fields $\boldsymbol{E}^\beta$ and $\boldsymbol{H}^\beta$ by the appropriate matrix $\widehat{M}_\beta$ in the proper frames.

From Eq. (16) describing Fresnel-Fizeau drag we can see that the small speed approximation $\beta \ll 1/n$ is not applicable to transformations between the $S_{mat}$-frame and the darkness $S_⊓$-frame and according to the exact velocity addition formula the entire phase velocity of the wave in $S_{mat}$ is due to Fresnel-Fizeau drag of its quasistatic high-k "reincarnation" in its proper darkness $S_⊓$-frame

$$\Delta v_{ph} = -v_{ph}$$

According to the nature of the fields' polarization of the high-k waves in their proper darkness $S_{\bar{n}}$-frame and the transformation Eq. (17), applied from $S_{\bar{n}}$ to $S_{mat}$ with $\beta = -1/n$, we get

$$\overline{\Delta g}_z^{\beta\hat{z}} = \bar{g}_{Min}^{\beta\hat{z}}, \qquad \bar{g}_{Abr}^{\beta\hat{z}} = 0, \quad U_{\beta\hat{z}} = 0$$

$$\frac{c\overline{\Delta g}_z}{U} = \frac{(1-\beta^2)}{-\beta} = n\left(1 - \frac{1}{n^2}\right) = nf$$

which is identical to Eq. (8).

This is an amazing result! We see that Fresnel-Fizeau drag coefficient when considered for transformation from the proper frame of the wave $S_{\bar{n}}$ disappears from the velocity addition formula but reappears in the expression for the hidden momentum. Note that in the latter case, we use not the index of refraction in the moving proper frame $S_{\bar{n}}$, where $n_\beta \to \infty$, but the index of refraction of the wave in the material rest frame $S_{mat}$. This shows that the longitudinal component of hidden momentum $\Delta g_z$ is ubiquitous for all waves with $n \neq 1$ and is relativistically related to the high-k dark field structure of waves in their proper darkness $S_{\bar{n}}$-frame.

6. Transverse hidden momentum and ray-wave tilt as differential aberration of ray and wave sources.

Now let us demonstrate that the ray-wave tilt and non-zero $\overline{\Delta g_\perp}$ can be understood as aberration analogous to the transverse Fresnel-Fizeau drag [59]. In a frame $S''$ moving in the direction perpendicular to the k-vector $\beta \cdot k = 0$, we can write

$$k'' = k - \gamma k_0 \beta$$

$$E'' = \gamma(E + \xi(\beta \cdot E)), \qquad H'' = \gamma(H + \xi(\beta \cdot H)), \qquad \xi = \left(\frac{1}{\gamma} - 1\right)\frac{1}{\beta^2}\beta + \frac{k}{k_0}$$

$$k'' \times g''_{Abr} = k'' \times \frac{1}{4\pi c}\text{Re}\{E''^* \times H''\}$$
$$= \frac{\gamma^2}{4\pi c}\text{Re}(\{(k + k_0 n^2 f\beta) \cdot H^*\}E'' - \{(k + k_0 n^2 f\beta) \cdot E\} H'')$$

From this identity we conclude that if $\beta$ is such that vector $k + k_0 n^2 f\beta$ is directed along the Abraham momentum in the laboratory frame, so that

$$(k + k_0 n^2 f\beta) \cdot H^* = (k + k_0 n^2 f\beta) \cdot E = 0$$

then in the frame $S''$ the Minkowski momentum is collinear with the Abraham momentum $k'' \times g''_{Abr} = 0$. If the Abraham momentum is tilted with respect to the Minkowski momentum at angle $\theta$, then the requirement of collinearity of $g''_{Abr}$ and $g''_{Min}$ is

$$\beta = \tan\theta /(nf) \tag{18}$$

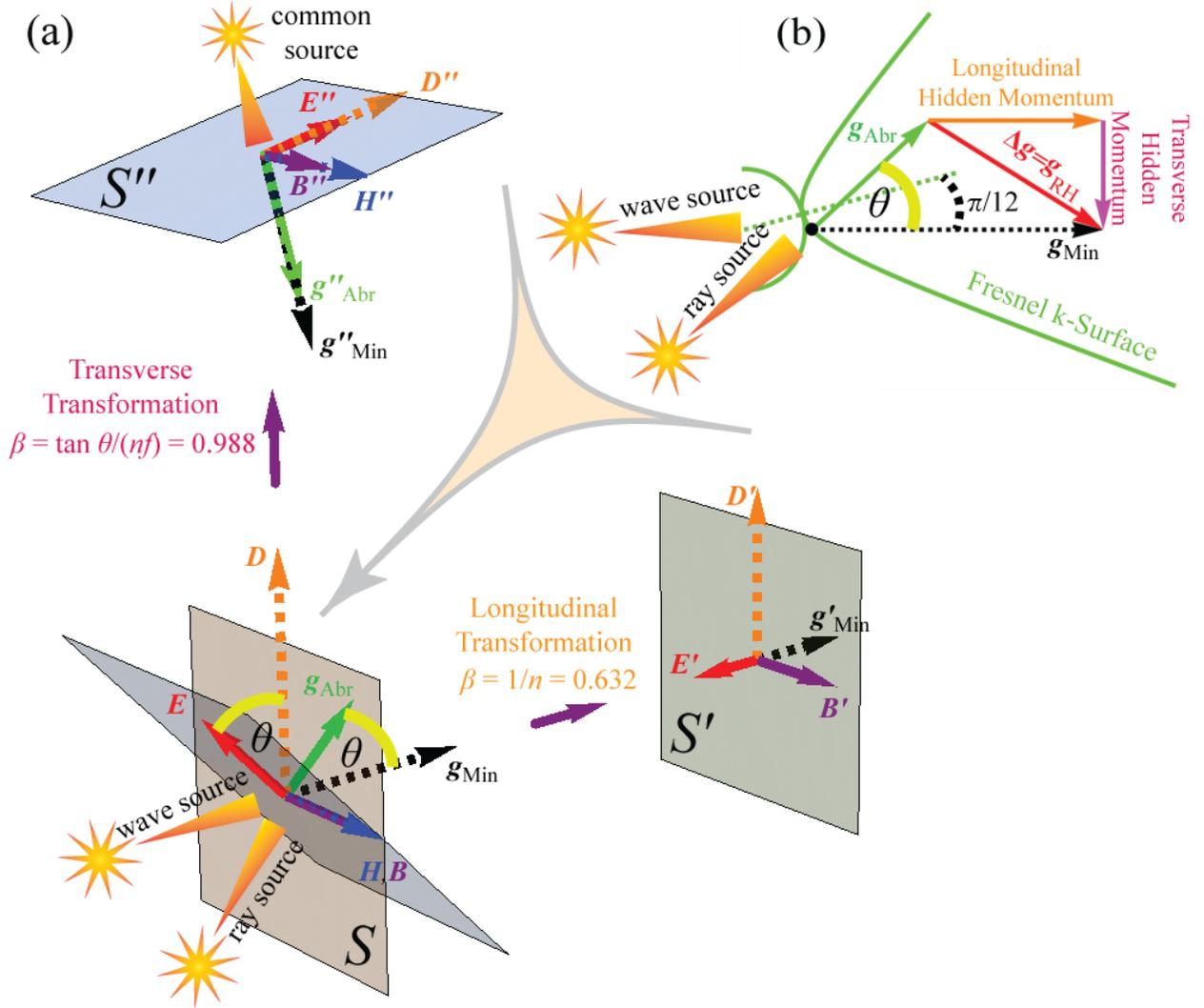

*Fig. 5. (a) The fields and apparent sources of rays and waves in frames $S = S_{mat}, S' = S_n, S''$ for a material with $\hat{\epsilon} = \mathrm{diag}\{2, 2 - 1\}$. In S frame the Abraham and Minkowski momenta have different longitudinal components and are at angle $\theta$ to each other, which corresponds to refraction of the wave because of Fresnel-Fizeau drag with respect to frame $S'$ and the divergence of apparent sources for rays and waves as the result of aberration from frame $S''$, where the ray and wave sources converge. (b) The hidden momentum breakdown in S frame, the ray and wave sources shown in reference to the Fresnel wave surface in k-space.*

It can be seen that the aberration is stronger for the rays than that for the phase if $n \neq 1$ and $f \neq 0$, since in the material rest frame $S_{mat}$ the direction $\boldsymbol{k}'' = \boldsymbol{k} - \gamma k_0 \boldsymbol{\beta}$ changes to $\boldsymbol{k}$ for the Minkowski momentum $\boldsymbol{g}_{Min}$, but all the way to $\boldsymbol{k} + k_0 n^2 f \boldsymbol{\beta}$ for the Abraham momentum $\boldsymbol{g}_{Abr}$. Like in the case of the longitudinal component of hidden momentum, the index of refraction $n$ and Fresnel-Fizeau drag coefficient $f$ in Eq. (18) correspond to the material rest frame $S_{mat}$. Note that the difference in aberration magnitude between rays and wave vectors leads to the appearance of the electric and magnetic bound charge waves as described above.

In Fig. 5 we schematically show the fields, the hidden momentum breakdown, and the apparent ray and wave sources in frames $S_{mat}, S' = S_\sqcap, S''$ for a nonmagnetic material with $\hat{\epsilon} = \text{diag}\{2,2-1\}$ which is free from magnetoelectric coupling depicted in Fig. 2. We consider a wave with $n = 1.582$ propagating in $xz$-plane at angle $\pi/12$ to the Fresnel wave surface axis (dashed green line) as shown in Fig. 2(b). Due to the nonmagnetic nature of this medium only the Roentgen hidden momentum $\boldsymbol{g}_{RH}$ is present. Transforming this wave from $S_{mat}$ to $S_\sqcap$ with $\beta = \frac{1}{n} = 0.632$ in the direction of the wave vector we find the wave becomes quasistatic high-k wave with $\boldsymbol{g}_{Abr} = 0$ with longitudinally polarized fields $\boldsymbol{E}, \boldsymbol{H}$. The transformation from $S_{mat}$ to $S_\sqcap$ can be better appreciated from Fig. 2(d), where the red curve corresponds to $S_{mat}$ and blue curve corresponds to $S_\sqcap$. We see how for increasing $\beta$ the intersection of the Fresnel wave surface with the z-axis happens at higher $n$, and reaches $n \to \infty$ for the blue curve with $\beta = \frac{1}{n} = 0.632$. Reverse transformation from $S_\sqcap$ into $S_{mat}$ leads to non-zero hidden momentum and refraction of the wave.

Transformation from $S_{mat}$ into $S''$ results in $\boldsymbol{g}_{Min}$ and $\boldsymbol{g}_{Abr}$ in the same direction. Transforming back from $S''$ into $S_{mat}$ leads to aberration of both $\boldsymbol{g}_{Min}$ and $\boldsymbol{g}_{Abr}$, however, this aberration is differential and results in different directions of $\boldsymbol{g}_{Min}$ and $\boldsymbol{g}_{Abr}$ in $S_{mat}$. This can be understood in terms of sources of waves and rays, which appear to coincide in $S''$, but are differentially aberrated and appear from different directions in $S_{mat}$. It should be noted that if the fields $\boldsymbol{E}$ and $\boldsymbol{H}$ in $S_{mat}$ are not linearly polarized the $S''$ frame is different for different times within the period of wave oscillations and at different locations.

In a general case of arbitrarily polarized waves, we turn to the "light" frames $S_\aleph$. As was described above, in $S_\aleph$ $\boldsymbol{n}_\beta \to -\boldsymbol{\hat{\beta}}$, and $n_\beta \to 1$, so that the waves appear as if they were under optical neutrality conditions [58]. Since $\gamma \gg 1$, the field components parallel $\boldsymbol{\beta}$ are the same as in $S_{mat}$, while the field components transverse to $\boldsymbol{\beta}$ are proportional to $\gamma$, $\boldsymbol{E}_t, \boldsymbol{H}_t, \boldsymbol{D}_t, \boldsymbol{B}_t \propto \gamma \to \infty$, which means that ray-wave tilt disappears at high $\gamma$. Note that even though $n \to 1$, the transverse field polarization is different from what it would be if the wave propagated in vacuum, meaning that in $S_\aleph$ frame the wave appears in an optical neutrality state and not as a pure vacuum field configuration. The optical neutrality here is in the sense that while the material parameters matrix $\widehat{M}$ is not an identity matrix, the eigenvalues of the index of refraction operator are $n \to \pm 1$.

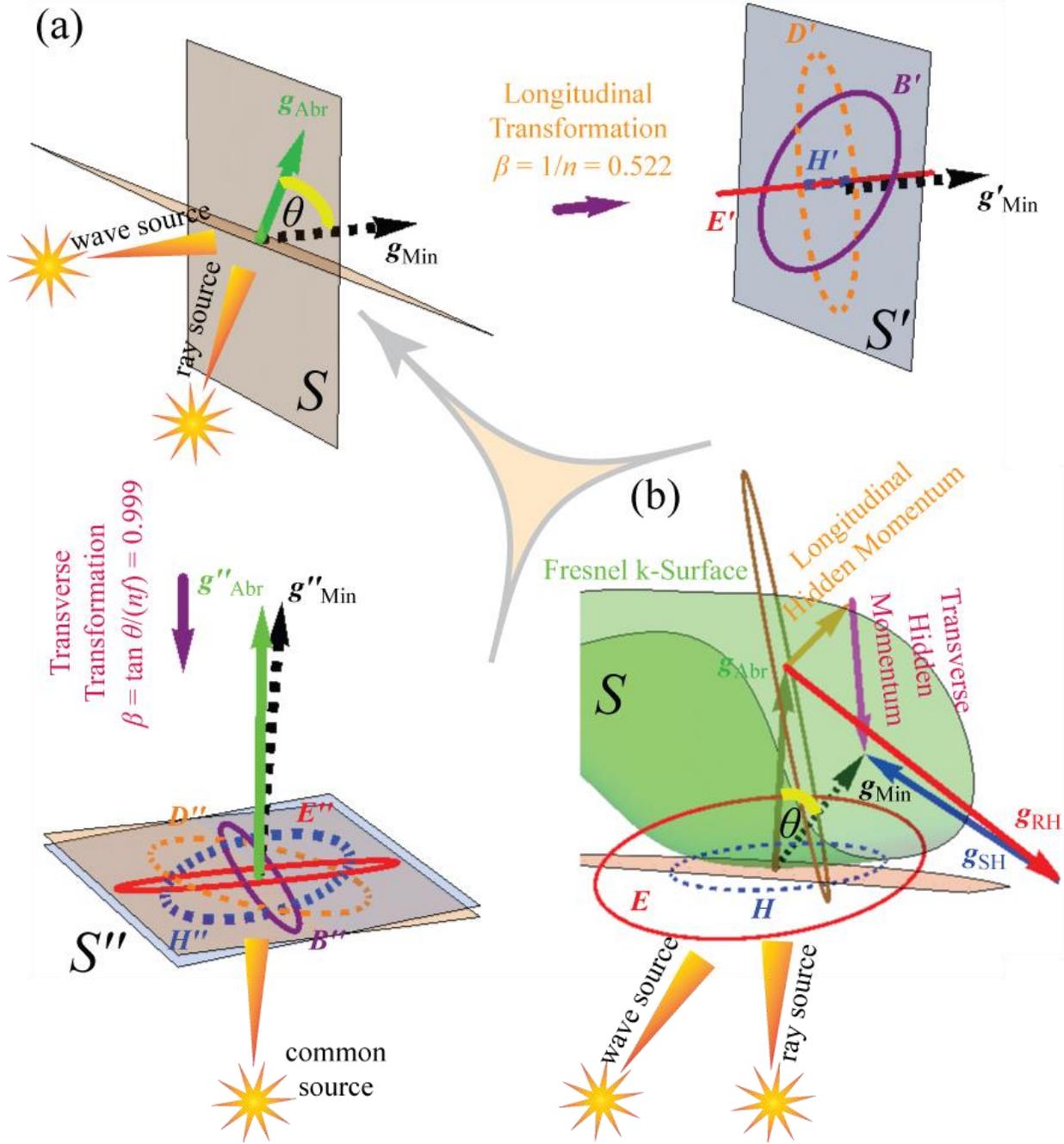

*Fig. 6. (a) The fields and apparent sources of rays and waves in frames $S = S_{mat}, S' = S_\Pi, S'' = S_\aleph$ for the same material as in Fig. 3. In S frame the Abraham and Minkowski momenta have different longitudinal components and are at angle θ to each other, which corresponds to refraction of the wave because of Fresnel-Fizeau drag with respect to frame S' and the divergence of apparent sources for rays and waves as the result of aberration from frame S'', where the ray and wave sources converge. (b) The hidden momentum breakdown in S frame, the ray and wave sources, and fields shown in reference to the Fresnel wave surface.*

In Fig. 6 we show the fields, the Roentgen and Shockley hidden momentum breakdown, and the apparent ray and wave sources in frames $S_{mat}, S' = S_ה, S'' = S_א$ for the material described in Fig. 3. We consider a wave with $n = 1.915$ propagating along the z-axis, i.e. $\boldsymbol{g}_{Min} \propto \hat{\boldsymbol{z}}$, as shown in Fig. 6(b) in reference to the Fresnel wave surface. In Fig. 6(b) we see that the fields $\boldsymbol{E}, \boldsymbol{H}$ in $S_{mat}$ are elliptically polarized and the instantaneous $\boldsymbol{g}_{Abr}$ follows the green ellipse. The time-average $\boldsymbol{g}_{Abr}$ points into the center of the green ellipse and is normal to the Fresnel wave surface. Transforming this wave from $S_{mat}$ to $S_ה$ with $\beta = \frac{1}{n} = 0.522$ in the direction of the wave vector we find that the wave becomes a quasistatic high-k wave with $\boldsymbol{g}_{Abr} = 0$ with longitudinally polarized fields $\boldsymbol{E}, \boldsymbol{H}$, while $\boldsymbol{D}, \boldsymbol{B}$ are still elliptically polarized. Reverse transformation into $S_{mat}$ leads to non-zero hidden momentum and refraction of the wave as described above. Transformation from $S_{mat}$ into $S''$ results in $\boldsymbol{g}_{Min}$ and $\boldsymbol{g}_{Abr}$ tending to the same direction due to $\boldsymbol{E}_t, \boldsymbol{H}_t, \boldsymbol{D}_t, \boldsymbol{B}_t \propto \gamma \to \infty$. Transforming back from $S''$ into $S_{mat}$ leads to aberration of both $\boldsymbol{g}_{Min}$ and $\boldsymbol{g}_{Abr}$, however, this aberration is differential and results in different directions of $\boldsymbol{g}_{Min}$ and $\boldsymbol{g}_{Abr}$ in $S_{mat}$. This can be understood in terms of sources of waves and rays, which appear to coincide in $S''$, but are differentially aberrated and appear from different directions in $S_{mat}$.

To conclude, we demonstrated that hidden electromagnetic momentum is intimately related to refraction and ray-wave tilt of electromagnetic waves in isotropy broken materials and can be understood in terms of Lorentz transformations between material rest frames $S_{mat}$, wave darkness proper frames $S_ה$, and optical neutrality light frames $S_א$ described in this manuscript. This is also related to the introduced in this manuscript fact that to moving observers the topological hyperbolic phases of optical materials may appear different from the phases in the material rest frames.